\documentclass[preprint,preprintnumbers,amsmath,amssymb]{revtex4}
\usepackage{booktabs}
\usepackage{mathrsfs}
\usepackage{epsfig}
\usepackage{graphicx}
\usepackage{dcolumn}
\usepackage{bm}
\usepackage{amsmath}
\usepackage{slashed}

\let\jnfont=\rm
\def\NPB#1,{{\jnfont Nucl.\ Phys.\ B }{\bf #1},}
\def\PLB#1,{{\jnfont Phys.\ Lett.\ B }{\bf #1},}
\def\EPJC#1,{{\jnfont Eur.\ Phys.\ Jour.\ C }{\bf #1},}
\def\PRD#1,{{\jnfont Phys.\ Rev.\ D }{\bf #1},}
\def\PRL#1,{{\jnfont Phys.\ Rev.\ Lett.\ }{\bf #1},}
\def\MPLA#1,{{\jnfont Mod.\ Phys.\ Lett.\ A }{\bf #1},}
\def\JPG#1,{{\jnfont J.\ Phys.\ G}{\bf #1},}
\def\CTP#1,{{\jnfont Commun.\ Theor.\ Phys.\ }{\bf #1},}
\def\ZPC#1,{{\jnfont Z.\ Phys.\ C }{\bf #1},}
\def\JHEP#1,{{\jnfont JHEP \ }{\bf #1},}
\def\Rv{\not{\hbox{\kern-1pt $R$}}}
\def\p{\not{\hbox{\kern-3pt $p$}}}

\begin{document}
\preprint{\parbox{1.2in}{\noindent }}

\title{SUSY Dark Matter In Light Of CDMS/XENON Limits}

\author{Jin  Min  Yang }

\affiliation{Key Laboratory of Frontiers in Theoretical Physics,
        Institute of Theoretical Physics, Academia Sinica,
        Beijing 100190, China
        \vspace*{1.5cm}}

\begin{abstract}
In this talk we briefly review the current CDMS/XENON constraints
on the neutralino dark matter in three popular supersymmetric
models: the minimal (MSSM), the next-to-minimal (NMSSM) and the nearly
minimal (nMSSM). The constraints from the dark matter relic density
and various collider experiments are also taken into account.
The conclusion is that for each model the current CDMS/XENON
limits can readily exclude a large part of the parameter space
allowed by other constraints and the future SuperCDMS or XENON100
can cover most of the allowed parameter space.
The implication for the Higgs search at the LHC is also discussed.
It is found that in the currently allowed parameter space the MSSM
charged Higgs boson is quite unlikely to be discovered at the LHC
while the neutral Higgs bosons $H$ and $A$ may be accessible
at the LHC in the parameter space with a large $\mu$ parameter.
\end{abstract}

\maketitle

\section{Introduction}
Low energy supersymmetry (SUSY) is the mainstream of new physics
beyond the standard model and often called the standard theory for
non-standard (new) physics. In addition to solve the fine-tuning
problem and ensure the unification of the gauge couplings, the
SUSY models (with $R$-parity) also have a pleasant byproduct-- the
cosmic dark matter. The lightest supersymmetric particle, which is
usually assumed to be the lightest neutralino (albeit gravitino
may equally be qualified), is a natural candidate for the cosmic
cold dark matter. Such a neutralino serves as a perfect WIMP
(Weakly Interacting Massive Particle) and is very likely within
the sensitivity reach of the dark matter direct detection
experiments like the CDMS \cite{cdms} or XENON \cite{xenon100-1}.

The current results from CDMS and XENON have set some limits on
the strength of the dark matter-nucleon scattering
\cite{cdms,xenon100-1}. These limits can be transferred to the
constraints on the SUSY parameter space under the assumption that
the cosmic dark matter is composed of the neutralino. Recently, in
our studies \cite{cao-1,cao-2} (for earlier studies, see, e.g.,
\cite{ellis}) a scan over the SUSY parameter space was performed
by considering the new CDMS/XENON limits together with the
constraints from the dark matter relic density and various
collider experiments like the search for the Higgs boson and super
particles, the precision electroweak measurements and the muon
anomalous magnetic moment. For comparison, several SUSY models
were taken in these studies, which are the minimal (MSSM), the
next-to-minimal (NMSSM) ) \cite{NMSSM1-13}
and the nearly minimal (nMSSM) \cite{xnMSSM1-5}.

In the meantime, the Large Hadron Collider at CERN will
intensively hunt for the SUSY particles including the neutralino
dark matter and the SUSY Higgs bosons (SUSY predicts at least five
Higgs bosons, i.e., in addition to the SM-like Higgs boson
$h_{SM}$, there are the neutral $H,A$ and the charged $H^\pm$).
Given a framework of SUSY model, the dark matter search
experiments like CDMS or XENON and the Higgs boson search at the
LHC will be correlated, and their interplay will allow for a deep
probe for SUSY. Focusing on the MSSM Higgs bosons, their
observability at the LHC is studied \cite{cao-2} in the parameter
space allowed by current dark matter and collider constraints.

Note that in this mini review we will not discuss the SUSY
explanation for the PAMELA cosmic ray anomaly \cite{pamela}, which
is subject to large uncertainty and could also be explained
astrophysically by pulsars. Usually, the MSSM, NMSSM or nMSSM are
hard to give a satisfactory explanation for such an anomaly (the
final states of the neutralino dark matter annihilation will give
excessive anti-protons or/and the large boost factor for the
annihilation rate is hard to generate from the elegant Sommerfeld
enhancement) \cite{kane,dm-nmssm} while some general singlet
extensions may give a plausible explanation
\cite{hooper,wang,yang} with some tension with the CDMS limits
\cite{tension}.

This review is organized as follows.
In Sec.\ref{sec2} we recapitulate the three SUSY models: the MSSM, NMSSM
and nMSSM.
Then we discuss the current CDMS/XENON constraints on the parameter
space of each model and their implication for the MSSM Higgs search
at the LHC in Sec.\ref{sec3}.
Finally, a summery is given in Sec. \ref{sec4}.

\section{SUSY models}
\label{sec2}
The MSSM has the minimal content of particles and thus is the most economical
realization of supersymmetry. So far this model has been most intensively
studied. However, this model suffers from the so-called $\mu$-problem and
the little hierarchy problem. Due to this reason, some singlet extensions
like the NMSSM and nMSSM have recently attracted much effort. In both the
NMSSM and the nMSSM the superpotential does not contain the $\mu$-term
which is generated by the non-zero vev of the newly introduced singlet
field. The little hierarchy can also be alleviated since the tree-level
upper bound on the SM-like Higgs boson is pushed up and the stop is not
necessarily so heavy to induce the large quantum effects for the Higgs
mass. Meanwhile, the LEP II lower bound of 114 GeV on the SM-like Higgs
boson is also somewhat relaxed since this Higgs boson has the singlet
mixture and its coupling with the $Z$-boson is weakened.
The superpotentials of these models are given by
\begin{eqnarray}
W_{\rm MSSM}& = & W_F+\mu \hat H_u\cdot\hat H_d,\\
W_{\rm NMSSM} & = & W_F + \lambda \hat{H}_u\cdot\hat{H}_d \hat{S}
                  +\frac{1}{3} \kappa  \hat{S}^3, \\
W_{\rm nMSSM} & = &  W_F + \lambda \hat{H}_u\cdot\hat{H}_d \hat{S}
                  + \xi_F M_n^2 \hat{S},
\label{superpotential}
\end{eqnarray}
where
$W_F= Y_u  \hat{Q}\cdot\hat{H}_u  \hat{U}
    -Y_d \hat{Q}\cdot\hat{H}_d \hat{D}
    -Y_e \hat{L}\cdot\hat{H}_d \hat{E}$
with $\hat{Q}$, $\hat{U}$ and $\hat{D}$ being the squark
superfields, and $\hat{L}$ and $\hat{E}$ being the slepton
superfields, $\hat{H}_u$ and $\hat{H}_d$ are the Higgs doublet
superfields,  $\lambda$, $\kappa$ and $\xi_F$ are dimensionless
coefficients, and $\mu$ and $M_n$ are parameters with mass
dimension.

So in the MSSM we have the SM-like Higgs boson $h_{SM}$, the heavy
neutral CP-even Higgs boson $H$ and CP-odd Higgs boson
$A$ plus the charged Higgs bosons $H^\pm$.
In the NMSSM and nMSSM, we have one more CP-even Higgs boson
and one more CP-odd Higgs boson. Because the lightest CP-even
Higgs boson and the lightest CP-odd Higgs boson may be dominated
by singlet components, they can escape the experimental constraints
and thus can be very light (sometimes called the light dark Higgs).

The neutralinos are the mixture of neutral gauginos and neutral
Higgsinos. While there are four neutralinos in the MSSM, the NMSSM
and nMSSM each has five neutralinos. The lightest one is usually
assumed to be the lightest superparticle (LSP). Due to the
$R$-parity conservation, this LSP is stable. Since it has weak
interaction and its mass is typically around the weak scale (a
light LSP at GeV scale may be also possible, see, e.g,
\cite{gev-lsp}), it is a perfect WIMP and serves as the
dark matter particle. The  neutralino mass matrices in these
models are given by
\begin{eqnarray}\label{mass-matrix1}
&
\left( \begin{array}{cccc}
M_1          & 0             & m_Zs_W s_b    & - m_Z s_W c_b   \\
0            & M_2           & -m_Z c_W s_b  & m_Z c_W c_b     \\
m_Zs_W s_b   & -m_Z s_W s_b  & 0             & -\mu            \\
-m_Z s_W c_b & -m_Z c_W c_b  &  -\mu         & 0               \\
\end{array} \right)
\end{eqnarray}
for the MSSM
\begin{eqnarray}\label{mass-matrix2}
\left( \begin{array}{ccccc}
M_1          & 0             & m_Zs_W s_b    & - m_Z s_W c_b  & 0 \\
0            & M_2           & -m_Z c_W s_b  & m_Z c_W c_b    & 0 \\
m_Zs_W s_b   & -m_Z s_W s_b  & 0             & -\mu           & -\lambda v c_b \\
-m_Z s_W c_b & -m_Z c_W c_b  &  -\mu         & 0              & - \lambda v s_b \\
0            & 0             &-\lambda v c_b &- \lambda v s_b & 2 \frac{\kappa}{\lambda}\mu
\end{array} \right)
\end{eqnarray}
for the NMSSM, and
\begin{eqnarray}
\left( \begin{array}{ccccc}
M_1          & 0             & m_Zs_W s_b    & - m_Z s_W c_b  & 0 \\
0            & M_2           & -m_Z c_W s_b  & m_Z c_W c_b    & 0 \\
m_Zs_W s_b   & -m_Z s_W s_b  & 0             & -\mu           & -\lambda v c_b \\
-m_Z s_W c_b & -m_Z c_W c_b  &  -\mu         & 0              & - \lambda v s_b \\
0            & 0             &-\lambda v c_b &- \lambda v s_b & 0
\end{array} \right)
\label{mass-matrix3}
\end{eqnarray}
for the nMSSM.
Here $M_1$ and $M_2$ are respectively $U(1)$ and $SU(2)$ gaugino masses,
$s_W=\sin \theta_W$, $c_W=\cos\theta_W$, $s_b=\sin\beta$ and $c_b=\cos\beta$
with $\tan \beta \equiv v_u/v_d$.

Note the property of the neutralino LSP is determined by its
components. In the MSSM with the GUT relation $M_2\simeq 2 M_1$,
it may be bino-like or Higgsino-like (without the GUT relation it
may also be wino-like). Since the LSP-nucleon scattering is
dominated by exchanging the Higgs bosons (for heavy squarks), a
bino-like neutralino scatters very weakly with the neucleon and
thus difficult to detect at CDMS/XENON. In the NMSSM, in the most
part of the parameter space the neutralino LSP is quite similar to
the case of MSSM \cite{cao-1}, but there exists some corner of the
parameter space in which the neutralino LSP is singlino-like and
thus can be very light \cite{gev-lsp}. In the nMSSM, due to the
zero diagonal element for the singlino in the mass matrix
Eq.(\ref{mass-matrix3}), the singlino is always very light and
thus the neutralino LSP is singlino-like. Therefore, the property
of the neutralino LSP in the nMSSM is quite different from the
MSSM or NMSSM.

\section{Numerical results and discussions}
\label{sec3}
In our study we consider the following
experimental constraints: (1) The direct bounds on sparticle and
Higgs masses from LEP and Tevatron experiments; (2) The LEP II search
for the Higgs bosons from various channels; (3) The LEP I and LEP II
constraints on the productions of neutralinos and charginos;
(4) The indirect constraints from the precision electroweak observables
and various $B$-decay and mixings;
(5) The  constraint from the muon anomalous magnetic moment,
for which we require the SUSY effects to account at $2 \sigma$ level;
(6)  The dark matter constraints from the WMAP relic density ($2\sigma$).
For the CDMS II/XENON100 direct detection exclusion limits ($90\%$ C.L.)
on the scattering cross section, we will display the results
with/without such limits. In addition to the above experimental limits,
we also consider the constraint from the stability of the Higgs potential,
which requires that the physical vacuum of the Higgs potential with
non-vanishing vevs of Higgs scalars should be lower than any local minima.

In our scan the soft breaking parameters are assumed to be below 1 TeV,
and the parameter $\lambda$ at weak scale is assumed to be less than
about 0.7 to ensure the perturbativity of the theory up to  the grand
unification scale.
Further, to reduce the number of the relevant soft parameters,
we assumed the so-called $m_h^{max}$ scenario with the soft
masses for the third generation squarks:
$M_{Q_3}=M_{U_3}=M_{D_3}=800$ GeV, and $X_t = A_t - \mu \cot \beta =
-1600$ GeV.

The parameter space surviving the above constraints is plotted in
Fig.1 of \cite{cao-1}. It shows that for each model the
CDMS-II/XENON100 limits can exclude a large part of the parameter
space allowed by other constraints and the future SuperCDMS (25
kg) or  XENON100 (6000 kg-days) can cover most of the allowed
parameter space. Still, some part of the allowed parameter space
is beyond the future sensitivity of the SuperCDMS or XENON100. To
totally cover the allowed parameter space of these SUSY models, a
much larger detector or larger exposure is needed. From this
figure we see that in the nMSSM the LSP can be as light as several
GeV while for the MSSM and NMSSM the LSP mass has a lower bound of
50 GeV which is from the assumed GUT relation $M_1 \simeq 0.5 M_2$
(plus the chargino lower bound of 103.5 GeV). Without such a GUT
relation, a lighter LSP is allowed. The similarity of the allowed
parameter space between the MSSM and NMSSM is due to the fact that
in the most part of the parameter space the LSP in both models is
bino-like. The peculiarity of the nMSSM parameter space is due to
the fact that the LSP is singlino-like in this model.

The property of the neutralino LSP in the allowed parameter space
is shown in Fig.2 of \cite{cao-1}. It shows that in case of
unobservation of dark matter the CDMS/XENON will push the
neutralino LSP more bino-like for the MSSM and NMSSM while more
singlino-like for the nMSSM. This can be understood easily because
a more bino-like or singlino-like LSP scatters more feebly with
the nucleon and more difficult to detect at the CDMS/XENON.

The charged Higgs boson mass range is shown Fig.7 of \cite{cao-1}. It shows that
in case of unobservation of dark matter the CDMS/XENON will push up the
charged Higgs boson mass for the MSSM and NMSSM. Then for the MSSM we studied
the charged  Higgs signal at the LHC $gg/gb \to t[b]H^+$ with $H^+$
subsequently decaying to $\tau^+ \nu_\tau$ \cite{cao-2}. As shown in Fig.4
of \cite{cao-2}, only a few survived points lie above the $5\sigma$ discovery
sensitivity obtained by the ATLAS collaboration for 30 fb$^{-1}$ integrated luminosity.
Thus the likelihood for discovering the charged Higgs boson at the LHC is very small.

For the MSSM neutral Higgs bosons $H$ and $A$ at the LHC, we
consider the channels $g g \to H(A)$ or $b\bar{b} H (A)$ with
$H(A)$ decaying to $\tau$ leptons. The results are shown in Fig.5
of \cite{cao-2}. It shows that in the allowed parameter space with
a large $\mu$ the neutral Higgs bosons $H$ and $A$ are quite
likely observable at the LHC. This can be understood from the fact
that for a larger  $\mu$ parameter the neutralino LSP is more
bino-like and scatters more feebly with the nucleon, which will
weaken the CDMS/XENON constraints.

Finally, for the light Higgs bosons in the NMSSM, some very recent
studies \cite{di-photon} showed that they may have
some interesting signals at the LHC, e.g., the enhanced di-photon
signal. Such enhanced di-photon signal may help to distinguish the
NMSSM from the MSSM.

\section{Conclusion}
\label{sec4} The current collider experiments and the dark matter
relic density measurement stringently constrained the SUSY
parameter space. The CDMS II and XENON100 limits further shrink
the allowed parameter space and pushed the neutralino dark matter
to be bino-like for the MSSM and NMSSM. The future sensitivity of
the SuperCDMS or XENON100 can cover most of the allowed  parameter
space for each model and in case of un-observation of dark matter
the neutralino dark matter will be more bino-like in the MSSM and
NMSSM while more singlino-like in the nMSSM. In the currently
allowed parameter space the MSSM charged Higgs boson is quite
unlikely to be discovered at the LHC while the neutral Higgs
bosons $H$ and $A$ may be accessible at the LHC in the parameter
space with a large $\mu$ parameter.

{\em Note added:~} When this manuscript is being prepared, more studies
about light SUSY dark matter appeared in the arXiv \cite{ldm-recent}.  

\section*{Acknowledgments}
This work was supported in part by the National Natural
Science Foundation of China under grant Nos. 10821504, 10725526 and 10635030.

\end{document}